\documentclass[sigconf]{acmart}
\copyrightyear{2026}
\acmYear{2026}
\setcopyright{cc}
\setcctype{by}
\acmConference[SIGIR '26] {Proceedings of the 49th International ACM SIGIR Conference on Research and Development in Information Retrieval}{July 20--24, 2026}{Melbourne, VIC, Australia.}
\acmBooktitle{Proceedings of the 49th International ACM SIGIR Conference on Research and Development in Information Retrieval (SIGIR '26), July 20--24, 2026, Melbourne, VIC, Australia}
\acmISBN{979-8-4007-2599-9/2026/07}
\acmDOI{10.1145/3805712.3808576}

\usepackage{subcaption}
\usepackage{multirow}
\usepackage{array}
\usepackage{colortbl}
\usepackage{makecell}
\usepackage{enumitem}

\usepackage{pifont}
\newcommand{\cmark}{\ding{51}} 
\newcommand{\xmark}{\ding{55}} 

\newcommand{\ie}{\emph{i.e., }}
\newcommand{\eg}{\emph{e.g., }}

\begin{document}
\title{iTIMO: An LLM-empowered Synthesis Dataset for Travel Itinerary Modification}

\author{Zhuoxuan~Huang}
\affiliation{%
  \institution{Central South University}
  \city{Changsha}
  \state{Hunan}
  \country{China}
}
\email{lilhzx@csu.edu.cn}

\author{Yunshan~Ma}
\affiliation{%
  \institution{Singapore Management University}
  \city{Singapore}
  \country{Singapore}
}
\email{ysma@smu.edu.sg}

\author{Hongyu~Zhang}
\authornote{Corresponding Authors.}
\affiliation{%
  \institution{Central South University}
  \city{Changsha}
  \state{Hunan}
  \country{China}
  }
\email{hyzhang@csu.edu.cn}

\author{Hua~Ma}
\affiliation{%
 \institution{Hunan Normal University}
 \city{Changsha}
 \state{Hunan}
 \country{China}
 }
 \email{huama@hunnu.edu.cn}

\author{Zhu~Sun}
\authornotemark[1]
\affiliation{%
  \institution{Singapore University of Technology and Design}
  \city{Singapore}
  \country{Singapore}
  }
\email{zhu_sun@sutd.edu.sg}

\renewcommand{\shortauthors}{Zhuoxuan Huang, Yunshan Ma, Hongyu Zhang, Hua Ma, Zhu Sun}

\begin{abstract}
   Addressing itinerary modification is crucial for enhancing the travel experience as it is a frequent requirement during traveling. However, existing research mainly focuses on fixed itinerary planning, leaving modification underexplored due to the scarcity of {\itshape need-to-modify} itinerary data. To bridge this gap, we formally define the itinerary modification task and propose a general pipeline to construct the corresponding dataset, namely iTIMO. This pipeline frames the generation of {\itshape need-to-modify} itinerary data as an intent-driven perturbation task. It instructs large language models to perturb real-world itineraries using three operations: REPLACE, ADD, and DELETE. Each perturbation is grounded in three intents: disruptions of popularity, spatial distance, and category diversity. Furthermore, hybrid metrics are proposed to ensure perturbation effectiveness. We conduct comprehensive benchmarking on iTIMO to analyze the capabilities and limitations of state-of-the-art LLMs. Overall, iTIMO provides a comprehensive testbed for the modification task, and empowers the evolution of traditional travel recommender systems into adaptive frameworks capable of handling dynamic travel needs. Dataset, code and supplementary materials are available at {\url{https://github.com/zelo2/iTIMO}}.

\end{abstract}

\ccsdesc[500]{Information systems~Information retrieval}
\ccsdesc[500]{Information systems~Evaluation of retrieval results}
\ccsdesc[500]{Information systems~Recommender systems}

\keywords{Travel Recommender Systems, Large Language Model, Travel Itinerary Modification, Synthetic Data Generation}

\maketitle

\section{Introduction}
    Travel Recommender Systems (TRSs) are instrumental in the tourism domain, enabling personalized itinerary planning in response to user inquiries \cite{chen2024deep}. Typically, given a user query specifying a departure point, destination, and various constraints ({\itshape e.g.,} time and budget), TRSs seek to construct a valid itinerary satisfying these constraints. Driven by their substantial commercial value, TRSs have garnered extensive attention from both academia and industry. Research on TRSs has continuously evolved, transitioning from early optimization-based approaches \cite{gasmi2024recommendation, yuan2024your, zhang2024collaborative} to deep learning models \cite{chen2023multi,chen2024geography, chen2023trip}. Recently, the advent of large language models (LLMs) has prompted researchers to investigate their potential in advancing TRSs \cite{xie2024travelplanner,chen2024travelagent} with a focus on the fixed itinerary planning.

\begin{figure}[t]
  \centering
  \includegraphics[height=0.30\textheight, keepaspectratio]{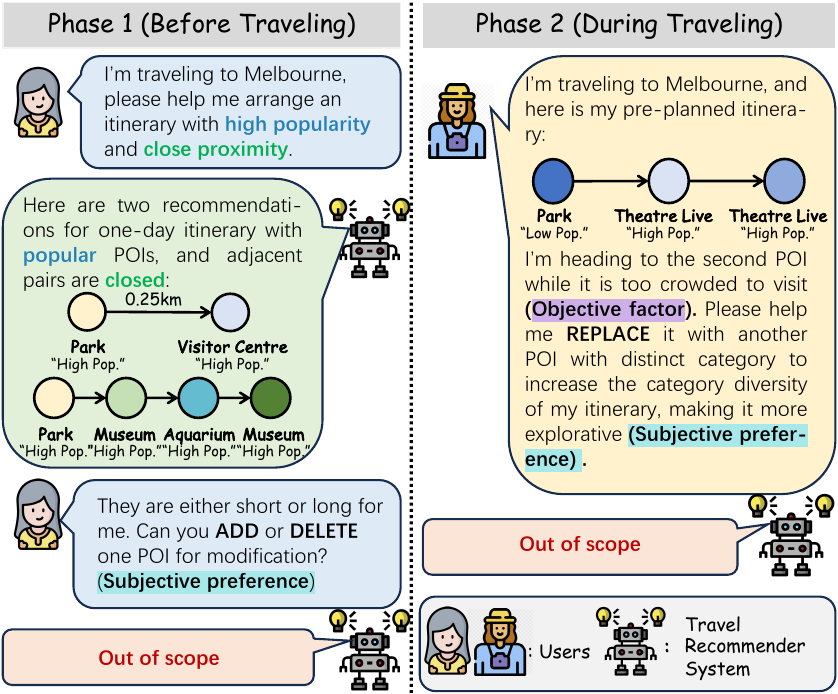}
  \caption{Illustrations of itinerary modification scenarios across two phases: before and during traveling. Users may proactively request to ADD, DELETE, or REPLACE POI for itinerary modification.}
  \Description{Two itinerary-modification scenarios before and during travel show user requests to add, delete, or replace points of interest.}
  \label{fig_1}
\end{figure}

However, this static generation paradigm contrasts with the dynamic nature of real-world travel, where itinerary modification represents an equally critical need. As illustrated in Figure~1, modification requests frequently occur across different phases, including both before and during traveling. Users often need to modify their itineraries by adding a new point-of-interest (POI) or replacing a undesired one due to subjective preferences or objective conditions. Consequently, the ability to address such modification requests is indispensable for ensuring a satisfactory travel experience. While current TRSs mainly focus on fixed itinerary planning, leaving modification underexplored. The primary reason for this gap is that public travel datasets \cite{xie2024travelplanner, banerjee2025synthtrips} contain real-world itineraries but lack the {\itshape need-to-modify} itineraries, which are essential for this task.

Generating {\itshape need-to-modify} itineraries is non-trivial. As shown in Figure~1, user inquiries stem from various intents ({\itshape e.g.,} seeking popular POIs in close proximity or pursuing category diversity for exploration) and operations ({\itshape e.g.,} adding, deleting, or replacing). This variability increases the complexity of constructing a comprehensive dataset for itinerary modification. Meanwhile, manual annotation is expensive and labor-intensive. Furthermore, relying on human validation risks introducing subjective bias that compromises reliability. In light of these obstacles, we aim to investigate: {\itshape Can we automatically generate need-to-modify itineraries to construct a diverse, scalable, and high-quality itinerary modification dataset?}

To achieve this goal, we formulate data construction as an intent-driven perturbation task grounded in real-world itineraries that implicitly encode user intents ({\itshape e.g.}, preferences regarding popularity). By strategically perturbing itineraries, we induce a divergence from the user's original intentions, thereby creating valid scenarios that necessitate modification and enabling the synthesis of realistic {\itshape need-to-modify} itineraries. This process requires managing complex dependencies within itineraries, such as maintaining spatial coherence or adhering to popularity preferences. To address this problem, we seek to leverage LLM, which is a promising approach given that its rich real-world knowledge enables perceiving various relationships \cite{xie2024travelplanner, gui2025hypertree, sun2024adaptive, guo2025enhancing, liu2025fine}.

Implementing this idea, we propose {\bfseries iTIMO}, an LLM-empowered synthetic dataset which comprises three sub-datasets for {\bfseries iTI}nerary {\bfseries MO}dification. 
{\bf First}, referring to prior research \cite{chen2023trip} that identifies popularity, spatial distance, and category diversity as key factors of user travel preference, we explicitly model these dimensions as distinct modification intents. We formally define the generation of {\it need-to-modify} itinerary data as an intent-driven perturbation task, executed via three operations: REPLACE, ADD, and DELETE~\footnote{ In this paper, we focus on these three operations that can satisfy multiple given intents, and leave other operations (\eg shuffle) as well as intents (\eg temporal constraints) as future work. Our proposed pipeline can be easily adapted to accommodate them.}. By applying these operations to itineraries under distinct intents, we effectively simulate diverse realistic scenarios where initial plans fail to meet user preferences, thereby establishing a comprehensive setting for the itinerary modification task. {\bf Second}, to ensure the generated itineraries satisfy the specified perturbation intents, we introduce hybrid evaluation metrics. They quantify the divergences between real-world itineraries and their perturbed versions from both macro- and micro-perspectives, providing measurable signals to facilitate {\itshape high-quality} perturbations. 
{\bf Third}, by integrating the proposed task formulation and evaluation metrics, we establish a generalized, LLM-empowered pipeline and apply it across three public travel datasets to construct iTIMO. This framework demonstrates {\it scalability} through its seamless applicability to public travel datasets. Moreover, in the proposed pipeline, we incorporate function calling and a memory module into an LLM to address the limitations of existing LLMs in itinerary perturbation ({\itshape cf.,} Section~5.2), further ensuring the {\itshape diversity} and {\itshape quality} of iTIMO dataset.

Besides the datasets, we conduct comprehensive benchmarking on iTIMO to systematically investigate the itinerary modification task. Specifically, we select various advanced baselines including eight base LLMs, and two large reasoning models (LRMs) to explore the boundaries of foundational model capabilities in handling such a novel task. Additionally, we explore the effects of retrieval-augmented generation (RAG) \cite{rag2020} and post-training strategies such as supervised fine-tuning (SFT) \cite{hu2022lora}, to provide a holistic view of effective paradigms for addressing the itinerary modification task. Our experiments yield several interesting insights. For example, small-scale models (\eg Qwen3-8B) with SFT can surprisingly outperform LRMs (\eg GPT o4-mini), and naively combining RAG with SFT does not always guarantee incremental improvements. Those findings shed light on promising future directions.

In summary, our contributions include:
\begin{itemize}[leftmargin=2em, topsep=2pt, itemsep=1pt, parsep=0pt, partopsep=0pt]
\item \textbf{A dataset for a novel task:} To the best of our knowledge, we are the first to propose the itinerary modification task and its corresponding iTIMO dataset to facilitate future research.
\item \textbf{A general pipeline for data generation:} We propose a general LLM-based pipeline to generate itinerary modification dataset based on real-world public datasets.
\item \textbf{Comprehensive benchmarking:} Extensive experiments on iTIMO datasets have unveiled meaningful insights and lead to several noteworthy and valuable directions for future research.
\end{itemize}

\section{Related Work}
\newcommand{\twoline}[2]{%
  \begingroup
  \renewcommand{\arraystretch}{0.85}
  \begin{tabular}[c]{@{}c@{}}#1\\[-0.35ex]#2\end{tabular}%
  \endgroup
}

\begin{table}[t]
  \centering
  \caption{Summary of existing travel datasets, where \cmark\ indicates the corresponding information is available, while \xmark\ indicates it is not.}
  \label{tab:dataset_compare}
  
\footnotesize
\setlength{\tabcolsep}{2.5pt}
\renewcommand{\arraystretch}{0.92}
\setlength{\extrarowheight}{-0.2pt}

  \begin{tabular*}{\columnwidth}{@{\extracolsep{\fill}}lccccc}
    \toprule
    \textbf{Dataset} & \textbf{User} & \textbf{POI}
      & \twoline{\textbf{Real-world}}{\textbf{itinerary}}
      & \twoline{\textbf{\textit{Need-to-modify}}}{\textbf{itinerary}}
      & \twoline{\textbf{Modification}}{\textbf{Intent}} \\
    \midrule
    \cite{wang2016improving, lim2015personalized, muntean2015learning}
      & \cmark & \cmark & \cmark & \xmark & \xmark \\
    \cite{xie2024travelplanner, wang2025triptailor, deng2025retail, ni2025tp}
      & \xmark & \cmark & \xmark & \xmark & \xmark \\
    \cite{shao2024chinatravel, qu2025tripscore}
      & \xmark & \cmark & \cmark & \xmark & \xmark \\
    \cite{tang2024itinera, shao2025personal, chaudhuri2025tripcraft}
      & \cmark & \cmark & \xmark & \xmark & \xmark \\
    \midrule
    \textbf{iTIMO (Ours)} & \cmark & \cmark & \cmark & \cmark & \cmark \\
    \bottomrule
  \end{tabular*}
\end{table}

\textbf{Benchmark Datasets for Itinerary Planning.}
Early studies \cite{lim2015personalized, muntean2015learning, wang2016improving} extract real-world itineraries from the YFCC100M dataset \cite{thomee2016yfcc100m} for itinerary planning. It contains geo-tagged photos from the Flickr platform. For instance, Wang et al. \cite{wang2016improving} proposed datasets including itineraries from Melbourne, and Muntean et al. \cite{muntean2015learning} collect real-world itinerary data from Florence, Pisa, and Roma. Recently, TravelPlanner \cite{xie2024travelplanner} constructs a comprehensive sandbox and synthetic user queries with multi-level constraints to justify the capability of LLM-empowered agents in handling itinerary planning task. After that, numerous efforts have been devoted to improve it in terms of dataset quality ({\itshape e.g.,} ChinaTravel \cite{shao2024chinatravel}, ITINERA \cite{tang2024itinera}, TripScore \cite{qu2025tripscore}, TripTailor \cite{wang2025triptailor}, RealTravel \cite{shao2025personal}, RETAIL \cite{deng2025retail}, and TP-RAG \cite{ni2025tp}) and evaluation metric ({\itshape e.g.,} TripScore \cite{qu2025tripscore}, Travel-Sim \cite{yangwide}, and TripCraft \cite{chaudhuri2025tripcraft}). Moreover, there are several studies focusing on generating personalized queries for travel destination recommendations \cite{banerjee2025synthtrips, wen2024elaborative}. These studies shed light on the capabilities of LLMs in itinerary planning. However, the scenario of itinerary modification is still underexplored. As summarized in Table~\ref{tab:dataset_compare}, existing datasets lack the essential data (\ie {\it need-to-modify} itinerary) to support itinerary modification study. This common but significant function remains a research gap, thereby limiting their core efficacy in handling dynamic changes and users' real-time needs, preventing them from becoming fully capable intelligent assistants.

\textbf{Itinerary Planning Methods.}
Itinerary planning task aims to provide personalized itineraries based on users' queries. Early studies model this task as an optimization problem, and solve it via classic algorithms \cite{gasmi2024recommendation, yuan2024your} or linear solvers \cite{zhang2024collaborative}. After that, reinforcement learning \cite{chen2023multi} and graph neural networks \cite{chen2024geography, chen2023trip} have been applied for personalized itinerary planning. More recently, inspired by TravelPlanner \cite{xie2024travelplanner}, numerous efforts have been devoted to improving LLMs' travel planning capabilities. For example, Hao et al. \cite{hao2025large} apply satisfiability modulo theory to the travel planning problem, which significantly improves LLMs' performance on the TravelPlanner benchmark. Gui et al. \cite{gui2025hypertree} and Zhe et al. \cite{zhe2025constraint} solve the planning problem from a hierarchical view, which decomposes complex reasoning chains into manageable sub-tasks and effectively handles diverse constraints. Aiming at the lack of personalization in existing LLM-based travel planning methods, Shao et al. \cite{shao2025personal} propose personal travel solver, which combines LLMs with numerical solvers to generate travel plans that satisfy both explicit constraints and implicit user preference. However, existing research focuses on generating static itineraries from user queries. This leaves a critical gap in itinerary modification, where existing plans need to be modified to address evolving user preferences or certain objective factors as illustrated in Fig.~1.

\section{Problem Definition}
    Given a set of real-world itineraries $\mathcal{I}$ composed of POIs from the POI set $\mathcal{P}$, an itinerary $i \in \mathcal{I}$ is a chronological sequence of POIs, where each POI is characterized by a tuple $\langle c, lat, long, pop \rangle$ denoting its category, geographic coordinates, and popularity. To systematically model modification scenarios, we generalize typical user requests into a set of operations $\mathcal{O}=\{o_{add},o_{replace},o_{delete}\}$, corresponding to adding a POI from the candidate set $\Phi_i = \mathcal{P} \setminus i$, replacing a POI with one from $\Phi_i$, and deleting a POI within $i$. Furthermore, grounded in established travel preferences ({\itshape i.e}., popularity, spatial distance, and diversity) \cite{chen2023trip}, we define the perturbation intent set $\mathcal{Z}=\{z_{pop},z_{dis},z_{div}\}$. These intents guide the generation of {\itshape need-to-modify} data by intentionally disrupting specific itinerary attributes. Formally, we define the two core tasks as follows:

{\bf Definition 1 (Itinerary Perturbation Task).} For each real-world itinerary $i \in \mathcal{I}$, the perturbation task aims to generate a {\itshape need-to-modify} itinerary $i^*$ by applying an operation $o \in \mathcal{O}$. The resulting $i^*$ is constrained to exhibit attribute disruptions aligned with the given intents $z \subseteq \mathcal{Z}$.

{\bf Definition 2 (Itinerary Modification Task).} Given a {\itshape need-to-modify} itinerary $i^*$, the itinerary modification task aims to revert it to the original itinerary $i$ through an operation $o \in \mathcal{O}$.
\section{{Proposed Metrics for} Itinerary Perturbation}
    
Itinerary perturbation quality is crucial for dataset construction. Since current methods lack mechanisms to verify if perturbations align with specific intents, we propose hybrid metrics to quantify intent fulfillment and ensure data reliability.

\subsection{{Measuring} Category Diversity Disruption}
For the intent $z_{div}$, given an itinerary $i$, we define there exists a disruption of category diversity once $div(i) \neq div(i^*)$, where $i^*$ is the perturbed itinerary based on $i$, $div(\cdot)$ is the function to compute category diversity:
\begin{equation}
{div}(i)=0\ \text{if }\#{unique}(c^i)=1;\quad
\#{unique}(c^i)/|i|\text{otherwise.}
\label{eq:cd}
\end{equation}
\par\noindent In Eq. (\ref{eq:cd}), $\#unique\left(c^{i}\right)$ denotes the number of unique POI categories in the itinerary $i$, $c^i$ is the {POI} categories within $i$, and $|i|$ represents the number of POIs in $i$. 

\subsection{{Measuring} Popularity Disruption}
For the intent $z_{pop}$, we define the popularity disruption from the macro and micro views. From the {\bfseries {macro}} view, popularity disruption refers to the popularity distribution shift between an itinerary and its perturbed version. To get the itinerary's popularity distribution, we categorize the popularity level of each POI within the itinerary into $\begin{Bmatrix}low,medium,high\end{Bmatrix}$ based on the visit frequency:
\begin{equation}
\mathcal{D}^{pop}(i)=\left\{\#pop_{i}^{l}\big/|i|,\ \#pop_{i}^{m}\big/|i|,\ \#pop_{i}^{h}\big/|i|\right\}
\label{eq:dpop},
\end{equation}
\par\noindent where $\mathcal{D}^{pop}(i)$ denotes the popularity distribution of itinerary $i$, $\#pop_{i}^{l}$, $\#pop_{i}^{m}$, and $\#pop_{i}^{h}$ denote the number of POIs with $low$, $medium$, and $high$ popularity level within $i$, respectively. 

Based on it, common metrics such as Jensen Shannon divergence (JSD) \cite{lin2002divergence},
total variation distance (TVD) \cite{lehmann2005testing}, or Hellinger distance (H) \cite{le2000asymptotics}
can be used to {measure} the macro distribution shifts.
We choose Hellinger distance as the metric, defined as follows:

\begin{equation}
\begin{split}
\mathrm{H}\!\left(\mathcal{D}^{pop}(i),\mathcal{D}^{pop}(i^{*})\right)
&=\frac{1}{\sqrt{2}}
\Biggl[
\sum_{k\in\{l,m,h\}}
\Bigl(
\sqrt{\#pop_{i}^{k}/|i|} \\
&\qquad\qquad
- \sqrt{\#pop_{i^{*}}^{k}/|i^{*}|}
\Bigr)^{2}
\Biggr]^{1/2}
\end{split}
\label{eq:hpop}
\end{equation}

\begin{figure}[t]
\centering
\includegraphics[width=1\columnwidth]{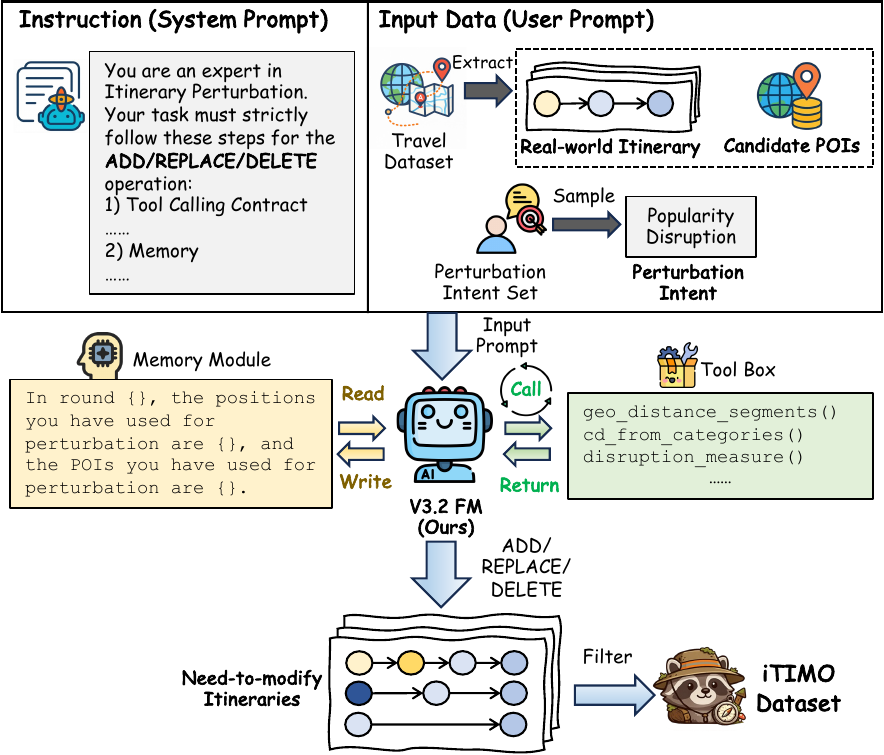} 
\caption{Overall pipeline of iTIMO dataset construction. The top part introduces the process of prompt construction ({\itshape cf.} Section 5.1). In the middle part propose V3.2 FM for high-quality itinerary perturbation ({\itshape cf.} Section 5.2). In the bottom part, we manually filter noise to construct iTIMO dataset ({\itshape cf.} Section 5.3).} 
\Description{Overall pipeline of the itinerary perturbation framework.}
\label{fig_2}
\end{figure}
\par\noindent Hellinger distance is selected for its superior sensitivity to extreme distribution shift cases. For instance, an itinerary $i$ contains 5 high-popularity, 5 medium-popularity, and 1 low-popularity POIs. When the low-popularity POI is deleted to generate $i^*$, the perturbation reflects a preference shift from tolerating to excluding low-popularity POIs. The resulting distributions are $\mathcal{D}^{pop}(i)=[5/11, 5/11, 1/11]$ and $\mathcal{D}^{pop}(i^*)=[1/2, 1/2, 0]$. Quantitatively, the Hellinger distance ($\text{H}=0.216$) yields a stronger signal than JSD ($0.0474$) or TVD ($0.0909$), thereby effectively capturing disruptions in our task.

From the {\bfseries micro} view, popularity disruption refers to the popularity ranking order shift. Given the itinerary $i$ with {3} high-popularity POIs, 1 medium-popularity POI and {3} low-popularity POIs, the perturbed itinerary $i^*$ is generated by deleting a low-popularity one. This perturbation reflects a preference shift moving from a preference for a mix of high and low-popularity POIs to a stronger preference for high-popularity POIs, which also satisfies $z^{pop}$. Although this preference change is semantically significant, its impact on the overall distribution is minimal, resulting in a low value of H ($0.069$). This underscores the insensitivity of distribution-based metrics to such nuanced micro-level preference shift, highlighting the need for complementary {measurement} methods.

{To address it, we adopt Kendall's $\tau_b \in [-1,1]$ to capture micro-level ranking-order shifts, following \cite{feng2018learning}. It is a pair-wise ranking metric that measures the ordinal agreement between two rankings. For example, $\tau_b=1$ indicates identical relative ordering, $\tau_b=0$ implies no ordinal correlation, while $\tau_b=-1$ indicates complete reversal. Formally, it is computed as follows:}
\begin{equation}
\tau_b=\frac{n_c-n_d}{\sqrt{(n_0-n_1)(n_0-n_2)}},
\label{eq:taub}
\end{equation}
\par\noindent{where $n_0$ is the total number of pairs formed by the three popularity levels (\ie $\{low, medium, high\}$). $n_c$ and $n_d$ denote the numbers of {concordant} and {discordant} pairs based on the frequency ranking of these levels. $n_1$ and $n_2$ represent the numbers of tied pairs within $i$ and $i^*$, respectively.}

In summary, we can incorporate the H value and $\tau_b$ to {measure} popularity disruption. Given an itinerary $i$, the perturbation results $i^*$ satisfies $z^{pop}$ if $\mathrm{H}(\mathcal{D}^{pop}(i),\mathcal{D}^{pop}(i^*))>\theta$ or $\tau_b<1$, where $\theta$ is the threshold. In this paper, we set $\theta=0.1$, and we theoretically prove the effectiveness of this value in the supplementary materials.

\subsection{{Measuring} Spatial Distance Disruption} 
    Analogously, we {measure} the disruption of spatial distance through the incorporation of H value and $\tau_b$. The spatial distance distribution of a given itinerary $i$ is obtained by:
\begin{equation}
\mathcal{D}^{dis}(i)=\left\{\#dis_{i}^{l}\big/|i|,\ \#dis_{i}^{m}\big/|i|,\ \#dis_{i}^{h}\big/|i|\right\}
\label{eq:ddis},
\end{equation}
\par\noindent where $\#dis_{i}^{l}$, $\#dis_{i}^{m}$, and $\#dis_{i}^{h}$ denote the number of POIs with $low$, $medium$, and $high$ distance level within $i$, respectively. 

To establish distance levels, we compute the distance between all successive POIs across all itineraries in the dataset via the Haversine formula~\cite{sinnott1984virtues}. Based on the statistical distribution of these collected distances, we categorize them into three distance levels: $low$, $medium$, and $high$.

\section{Pipeline for iTIMO Dataset Construction}
    This section details the integration of our proposed metrics { in Section 4} to generate high-quality itinerary perturbations for iTIMO construction. The overall pipeline is illustrated in Figure~\ref{fig_2}. 
\begin{table}[!t]
  \centering
  \caption{Comparison between DeepSeek models and our V3.2 FM on different perturbation operations. $\uparrow$ ($\downarrow$) indicates higher (lower) is better.}

  \label{tab:poi-div-pert-acc-compact}

  \scriptsize
  \setlength{\tabcolsep}{2pt}
  \renewcommand{\arraystretch}{0.92}

  \begin{tabular*}{\columnwidth}{@{\extracolsep{\fill}}c|cccccc@{}}
    \toprule
    \textbf{Operation} & \textbf{Model}
    & \makecell[c]{\textbf{POI}\\\textbf{Div.}$\uparrow$}
    & \makecell[c]{\textbf{Pert.}\\\textbf{Acc.}$\uparrow$}
    & \makecell[c]{\textbf{Avg.}\\\textbf{time}$\downarrow$}
    & \makecell[c]{\textbf{Avg.}\\\textbf{tokens}$\downarrow$}
    & \makecell[c]{\textbf{Avg. cost}\\\textbf{(\$)}$\downarrow$} \\
    \midrule

    \multirow{3}{*}{ADD}
      & V3.2                      & 0.15 & 0.70 &  27.3s &  3925.7  & 0.0024 \\
      & R3.2                      & 0.54 & 0.94 & 884.3s & 21998.9 & 0.0289 \\
      & {\bfseries V3.2 FM (Ours)}& \textbf{0.84} & \textbf{1.00} & 102.4s & 10470.1 & 0.0042 \\
    \midrule

    \multirow{3}{*}{DELETE}
      & V3.2                      & 0.68 & 0.62 &  23.5s &  2023.9  & 0.0019 \\
      & R3.2                      & 0.66 & 0.98 & 474.5s & 14171.5 & 0.0186 \\
      & {\bfseries V3.2 FM (Ours)}& \textbf{0.70} & \textbf{1.00} &  96.8s & 12595.6 & 0.0048 \\
    \midrule

    \multirow{3}{*}{REPLACE}
      & V3.2                      & 0.19 & 0.68 &  25.7s &  3735.7  & 0.0024 \\
      & R3.2                      & 0.72 & 0.98 & 590.0s & 18998.6 & 0.0249 \\
      & {\bfseries V3.2 FM (Ours)}& \textbf{0.74} & \textbf{1.00} & 100.3s & 13246.3 & 0.0049 \\
    \bottomrule
  \end{tabular*}
\end{table}

\subsection{Prompt Template Design} 
For each operation $o \in \mathcal{O}$, we design a prompt template $\pi_o$. The system prompt establishes foundational task context, incorporating our hybrid evaluation metrics to provide semantic grounding for high-quality perturbations. The user prompt then provides the contextualized input, including the original itinerary $i \in \mathcal{I}$, candidate POIs $\Phi_{i}$, and target perturbation intents $z \subseteq \mathcal{Z}$. To ensure diverse perturbation intents, we constrain the number of intents $|z|$ by uniformly sampling from $\{1, 2, 3\}$. Formally, the prompt-based perturbation process for each operation $o$ is formulated as:
\begin{equation}
    i^*=f_\Theta\left(i,\Phi_{i},z\mid \pi_o\right),
\end{equation}
\par\noindent where $f_\Theta(\cdot)$ denotes the LLM whose parameters are denoted by $\Theta$.

\definecolor{POIPurple}{RGB}{55, 20, 90}     
\definecolor{PertPink}{RGB}{236, 175, 188}   

\DeclareRobustCommand{\legendbox}[1]{%
  \textcolor{#1}{\rule{0.9em}{0.9em}}%
}

\begin{figure}[h]
  \centering
  \captionsetup[subfigure]{font=scriptsize, skip=0pt, labelfont=bf, textfont=bf}

  \setlength{\tabcolsep}{2pt}

  \begingroup
  \setlength{\parskip}{0pt} 

  \begin{tabular}{@{}ccc@{}}
    \begin{subfigure}[b]{0.31\columnwidth}
      \centering
      \includegraphics[width=\linewidth,clip,trim=18pt 22pt 6pt 10pt]{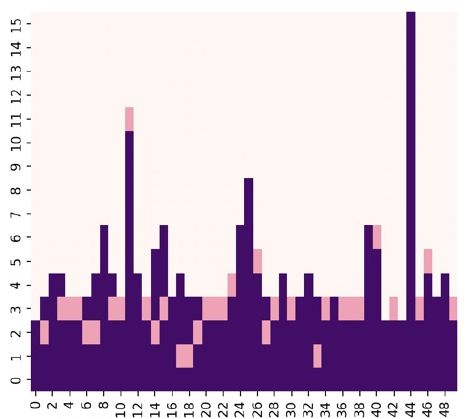}
      \caption{V3.2 / ADD}
    \end{subfigure}
    &
    \begin{subfigure}[b]{0.31\columnwidth}
      \centering
      \includegraphics[width=\linewidth,clip,trim=18pt 22pt 6pt 10pt]{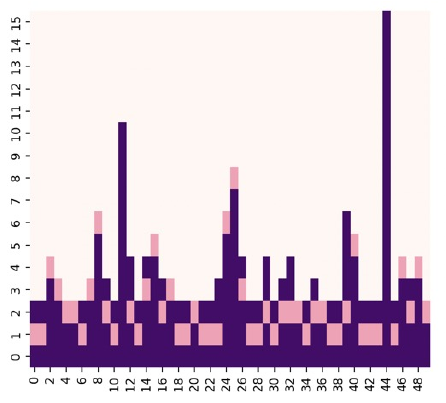}
      \caption{V3.2 / DELETE}
    \end{subfigure}
    &
    \begin{subfigure}[b]{0.31\columnwidth}
      \centering
      \includegraphics[width=\linewidth,clip,trim=18pt 22pt 6pt 10pt]{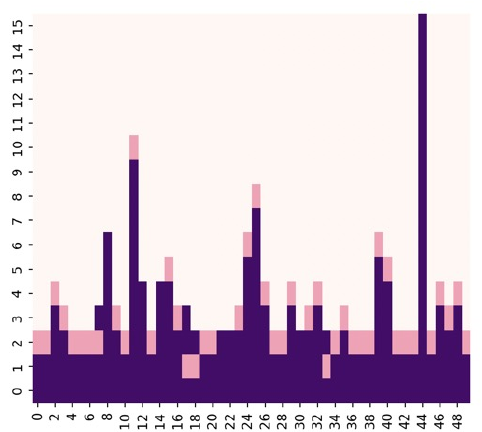}
      \caption{V3.2 / REPLACE}
    \end{subfigure}
    \\[-1mm]

    \begin{subfigure}[b]{0.31\columnwidth}
      \centering
      \includegraphics[width=\linewidth,clip,trim=18pt 22pt 6pt 10pt]{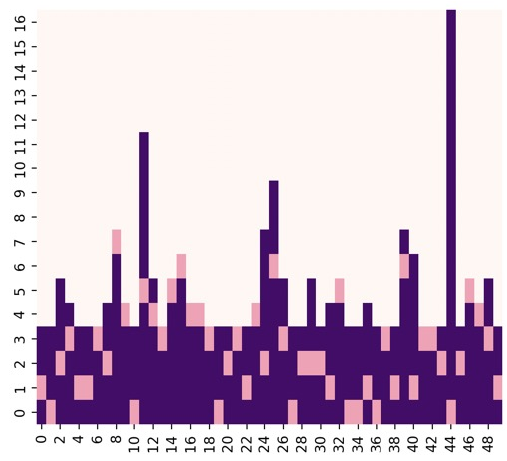}
      \caption{V3.2 FM / ADD}
    \end{subfigure}
    &
    \begin{subfigure}[b]{0.31\columnwidth}
      \centering
      \includegraphics[width=\linewidth,clip,trim=18pt 22pt 6pt 10pt]{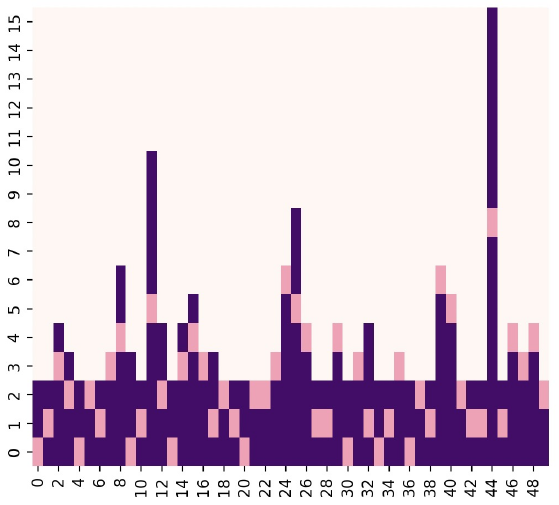}
      \caption{V3.2 FM / DELETE}
    \end{subfigure}
    &
    \begin{subfigure}[b]{0.31\columnwidth}
      \centering
      \includegraphics[width=\linewidth,clip,trim=18pt 22pt 6pt 10pt]{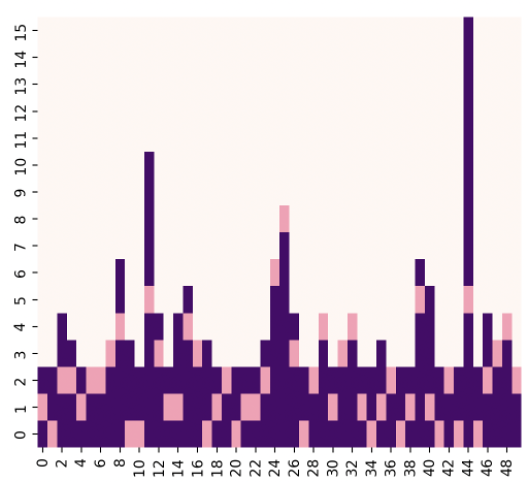}
      \caption{V3.2 FM / REPLACE}
    \end{subfigure}
  \end{tabular}

  \endgroup
  \caption{Visualization of perturbation positions for V3.2 and our V3.2 FM, where each column corresponds to one itinerary, \legendbox{POIPurple} indicates an existing POI, and \legendbox{PertPink} indicates a perturbation position.}
  \Description{Six heatmaps compare perturbation positions for V3.2 and V3.2 FM across add, delete, and replace operations. Dark purple marks existing POIs and pink marks perturbation positions.}
  \label{fig:mel-posdiv-v32-2x3}
\end{figure}

\subsection{Itinerary Perturbation Method}
We conduct a preliminary study to investigate the effectiveness of LLMs in the itinerary perturbation task. Our experimental datasets are constructed by sampling 50 itineraries from each of the three public travel datasets: Toronto \cite{lim2015personalized}, Melbourne \cite{wang2016improving}, and Florence \cite{muntean2015learning}. We utilize advanced DeepSeek models \cite{liu2024deepseek} as baselines under two settings: a thinking mode (R3.2) and a non-thinking mode (V3.2). Due to space constraints and consistent trends across datasets, we report results exclusively for Melbourne. Experimental results are presented in Table~\ref{tab:poi-div-pert-acc-compact}, {where POI Div. is defined as the ratio of the number of unique  POIs selected for perturbation to the total number of samples, reflecting the perturbation diversity}, and Pert. Acc. measures perturbation accuracy by assessing the alignment between perturbation results and intents.

From Table~\ref{tab:poi-div-pert-acc-compact}, R3.2 outperforms V3.2 in both diversity and accuracy, demonstrating the priority of reasoning capability in itinerary perturbation task. However, the long reasoning context of R3.2 leads to inefficiencies and high token consumption, making the model computationally and financially expensive to deploy. To better balance the trade-off between performance and efficiency, we build upon V3.2 to propose V3.2 with {\bfseries f}unction calling (FC) and {\bfseries m}emory module, termed as {\bfseries V3.2 FM}. They are introduced to address the accuracy and diversity issues, respectively. 

Regarding the {\bfseries accuracy issue}, analysis of V3.2 and R3.2 failure cases attributes it to computational limitations. For example, scrutinizing the R3.2's chain-of-thought reveals the model's reliance on an estimation strategy for spatial distance computation, explicitly stating {\itshape "Since I am a language model, I will approximate it.}" This approximation may misclassify short spatial distances, erroneously categorizing low-level distances into the medium-level category. To remedy this, we integrate FC to replace error-prone textual estimation with precise computation. As illustrated in Figure~{\ref{fig_2}}, we build a tool box to enable the model to access exact numerical values via FC, effectively eliminating computational hallucinations caused by approximation. Experimental results reported in Table~\ref{tab:poi-div-pert-acc-compact} demonstrate that our V3.2 FM achieves optimal perturbation accuracy across all operations, while simultaneously reducing inference time and token consumption compared to the R3.2 reasoning model.

Regarding the {\bfseries diversity issue}, beyond the limited POI div. reported in Table~\ref{tab:poi-div-pert-acc-compact}, visualizations in Figure~{\ref{fig:mel-posdiv-v32-2x3}}~(a)-(c) also confirm that V3.2 suffers from a severe positional bias, failing to explore the full range of modification points. To address this deficiency, we introduce a memory module, inspired by \cite{cai2025agentic}. As shown in Figure~{\ref{fig_2}}, this module records the history of selected POIs and perturbation positions. By injecting this memory into the input, we effectively discourage the LLM from repeating previous modification actions. Experimental results shown in both Table~\ref{tab:poi-div-pert-acc-compact} and Figure~{\ref{fig:mel-posdiv-v32-2x3}}~(d)-(f) verify the effectiveness of the proposed memory module.

{Notably, our pipeline is highly generalizable, since it can be seamlessly applied to any travel dataset with standard POI features for expansion, and it is also extensible in terms of perturbation logic. This design facilitates the incorporation of new perturbation operations or intents, enabling the pipeline to evolve and adapt to simulating more complex itinerary modification scenarios.}
\newcommand{\pdd}[1]{\phantom{00}#1}
\newcommand{\pddd}[1]{\phantom{0}#1}
\begin{table}[t]
  \centering
  \caption{Statistics information of iTIMO dataset, where $\lvert z\rvert$ denotes the number of intents behind one perturbed itinerary.}
  \label{tab:itimo-stats}
  \footnotesize
  \setlength{\tabcolsep}{3pt}
  \begin{tabular*}{\columnwidth}{@{\extracolsep{\fill}}c|ccc|ccc|ccc}
    \toprule
    \multirow{2}{*}[-0.4ex]{\textbf{Metric}}
      & \multicolumn{3}{c|}{\textbf{iTIMO-Toronto}}
      & \multicolumn{3}{c|}{\textbf{iTIMO-Melbourne}}
      & \multicolumn{3}{c}{\textbf{iTIMO-Florence}} \\
    \cmidrule(lr){2-4}\cmidrule(lr){5-7}\cmidrule(lr){8-10}
      & {\scriptsize\textbf{DEL.}} & {\scriptsize\textbf{ADD}} & {\scriptsize\textbf{REP.}}
      & {\scriptsize\textbf{DEL.}} & {\scriptsize\textbf{ADD}} & {\scriptsize\textbf{REP.}}
      & {\scriptsize\textbf{DEL.}} & {\scriptsize\textbf{ADD}} & {\scriptsize\textbf{REP.}} \\
    \midrule
      \multicolumn{1}{l|}{\# POIs}
    & \pdd{26} & \pdd{26} & \pdd{26}
    & \pddd{160} & \pddd{160} & \pddd{160}
    & \pddd{862} & \pddd{862} & \pddd{862} \\
  \multicolumn{1}{l|}{\# Itineraries}
    & \pddd{334} & \pddd{321} & \pddd{333}
    & \pddd{401} & \pddd{375} & \pddd{398}
    & 4439 & 4256 & 4381 \\
  \multicolumn{1}{l|}{Avg.\ length}
    & \pddd{4.71} & \pddd{2.74} & \pddd{3.71}
    & \pddd{5.61} & \pddd{3.69} & \pddd{4.57}
    & 10.27 & \pddd{8.19} & \pddd{9.13} \\
  \midrule
  \multicolumn{1}{l|}{\# ($\lvert z\rvert = 1$)}
    & \pdd{23} & \pddd{105} & \pdd{44}
    & \pdd{35} & \pddd{105} & \pdd{91}
    & \pddd{845} & 1545 & 1255 \\
  \multicolumn{1}{l|}{\# ($\lvert z\rvert = 2$)}
    & \pddd{110} & \pddd{150} & \pddd{182}
    & \pddd{192} & \pddd{201} & \pddd{195}
    & 2182 & 1992 & 2255 \\
  \multicolumn{1}{l|}{\# ($\lvert z\rvert = 3$)}
    & \pddd{201} & \pdd{66} & \pddd{107}
    & \pddd{174} & \pdd{69} & \pddd{112}
    & 1412 & \pddd{719} & \pddd{871} \\
  \multicolumn{1}{l|}{\# ${z}_{div}$}
    & \pddd{253} & \pddd{166} & \pddd{156}
    & \pddd{253} & \pddd{171} & \pddd{155}
    & 3229 & 2857 & 1909 \\
  \multicolumn{1}{l|}{\# ${z}_{dis}$}
    & \pddd{317} & \pddd{185} & \pddd{289}
    & \pddd{352} & \pddd{293} & \pddd{326}
    & 3429 & 2409 & 3202 \\
  \multicolumn{1}{l|}{\# ${z}_{pop}$}
    & \pddd{276} & \pddd{252} & \pddd{284}
    & \pddd{336} & \pddd{250} & \pddd{336}
    & 2787 & 2420 & 3267 \\
    \bottomrule
  \end{tabular*}
\end{table}

\subsection{iTIMO Dataset}
The iTIMO dataset, comprising of three sub-datasets, is constructed by utilizing { our} V3.2 FM to perturb real-world itineraries derived from three public travel datasets: Toronto \cite{lim2015personalized}, Melbourne \cite{wang2016improving}, and Florence \cite{muntean2015learning}. This selection ensures coverage across diverse geographical and cultural contexts. After that, we manually filter errors and noise to guarantee the data quality. The statistical information of iTIMO are summarized in Table~{\ref{tab:itimo-stats}}.

\section{Experiment}

\definecolor{TypeLLM}{RGB}{198,219,239}  
\definecolor{TypeLRM}{RGB}{232,220,244}  
\definecolor{TypeSFT}{RGB}{252,232,200}  

\newcommand{\TypeLLM}{\rowcolor{TypeLLM}}
\newcommand{\TypeLRM}{\rowcolor{TypeLRM}}
\newcommand{\TypeSFT}{\rowcolor{TypeSFT}}

\newcommand{\legendrect}[1]{%
  \textcolor{#1}{\rule[0.1ex]{1.4em}{1.0ex}}%
}

\newcommand{\ColMax}[1]{\textbf{#1}}
\newcommand{\SetBest}[1]{\underline{\mbox{#1}}}
\newcommand{\BestBoth}[1]{\textbf{{\mbox{#1}}}}

\newsavebox{\MainTabBox}

\begin{table*}[t]
  \centering
  \caption{Itinerary modification performance across datasets, operations, LLMs, and RAG settings, where \legendrect{TypeLLM} denotes base LLMs, \legendrect{TypeLRM} denotes LRMs, \legendrect{TypeSFT} denotes LLMs with SFT, \texttt{bold} indicates the best performance in each column, and $\uparrow$ indicates higher is better. For base LLMs, we only keep zero-shot and the best setting selected by the average of Mod across all datasets and operations due to limited space.}
  \label{tab:main_result}

  \scriptsize
  \setlength{\tabcolsep}{2pt}
\renewcommand{\arraystretch}{0.85}

\sbox{\MainTabBox}{%
\resizebox{\textwidth}{!}{%
\begin{tabular}{@{}l|*{6}{c}|*{6}{c}|*{6}{c}@{}}
  \toprule
  & \multicolumn{6}{c|}{\textbf{iTIMO-Melbourne}}
  & \multicolumn{6}{c|}{\textbf{iTIMO-Toronto}}
  & \multicolumn{6}{c}{\textbf{iTIMO-Florence}} \\
  \cmidrule(lr){2-7}\cmidrule(lr){8-13}\cmidrule(lr){14-19}
  \multicolumn{1}{c|}{\textbf{Methods}} &
    \multicolumn{2}{c}{\textbf{ADD}} &
    \multicolumn{2}{c}{\textbf{REPLACE}} &
    \multicolumn{2}{c|}{\textbf{DELETE}} &
    \multicolumn{2}{c}{\textbf{ADD}} &
    \multicolumn{2}{c}{\textbf{REPLACE}} &
    \multicolumn{2}{c|}{\textbf{DELETE}} &
    \multicolumn{2}{c}{\textbf{ADD}} &
    \multicolumn{2}{c}{\textbf{REPLACE}} &
    \multicolumn{2}{c}{\textbf{DELETE}} \\
  \cmidrule(lr){2-19}
  &
    \textbf{Mod}${\uparrow}$ & \textbf{APR}${\uparrow}$ &
    \textbf{Mod}${\uparrow}$ & \textbf{APR}${\uparrow}$ &
    \textbf{Mod}${\uparrow}$ & \textbf{APR}${\uparrow}$ &
    \textbf{Mod}${\uparrow}$ & \textbf{APR}${\uparrow}$ &
    \textbf{Mod}${\uparrow}$ & \textbf{APR}${\uparrow}$ &
    \textbf{Mod}${\uparrow}$ & \textbf{APR}${\uparrow}$ &
    \textbf{Mod}${\uparrow}$ & \textbf{APR}${\uparrow}$ &
    \textbf{Mod}${\uparrow}$ & \textbf{APR}${\uparrow}$ &
    \textbf{Mod}${\uparrow}$ & \textbf{APR}${\uparrow}$ \\
  \midrule


\TypeLLM{Gemma3-4b}
  & 0.092 & 0.224 & 0.161 & 0.123 & 0.333 & 0.395
  & 0.077 & 0.215 & 0.119 & 0.149 & 0.118 & 0.309
  & 0.072 & 0.343 & 0.099 & 0.128 & 0.151 & 0.249 \\
\quad+RAG (Qwen3-8b-emd)
  & {0.132} & {0.421} & {0.161} & {0.210} & {0.358} & {0.469}
  & {0.139} & {0.369} & {0.134} & {0.194} & {0.397} & {0.500}
  & {0.073} & {0.358} & {0.079} & {0.125} & {0.152} & {0.253} \\

\TypeLLM{Llama-3.1-8b-Instruct}
  & 0.092 & 0.289 & 0.074 & 0.136 & 0.235 & 0.284
  & 0.077 & 0.308 & 0.075 & 0.104 & 0.176 & 0.368
  & 0.057 & 0.308 & 0.073 & 0.129 & 0.183 & 0.260 \\
\quad+RAG (KaLM-Gemma3-12b-emd)
  & {0.118} & {0.316} & {0.062} & {0.099} & {0.358} & {0.383}
  & {0.108} & {0.308} & {0.104} & {0.134} & {0.235} & {0.426}
  & {0.067} & {0.356} & {0.113} & {0.113} & {0.216} & {0.300} \\

\TypeLLM{Qwen3-8b}
  & 0.118 & 0.382 & 0.025 & 0.049 & 0.346 & 0.370
  & 0.077 & 0.385 & 0.119 & 0.194 & 0.324 & 0.515
  & 0.085 & 0.397 & 0.113 & 0.165 & 0.242 & 0.329 \\
\quad+RAG (GPT-text-emd-3-large)
  & {0.158} & {0.408} & {0.111} & {0.198} & {0.469} & {0.518}
  & {0.108} & {0.323} & {0.164} & {0.224} & {0.456} & {0.559}
  & {0.103} & {0.438} & {0.139} & {0.194} & {0.255} & {0.344} \\

\TypeLLM{Phi4-15b}
  & 0.132 & 0.355 & 0.086 & 0.123 & 0.185 & 0.185
  & 0.031 & 0.169 & 0.060 & 0.134 & 0.118 & 0.338
  & 0.055 & 0.320 & 0.062 & 0.127 & 0.165 & 0.244 \\
\quad+RAG (Qwen3-8b-emd)
  & {0.145} & {0.289} & {0.099} & {0.161} & {0.272} & {0.296}
  & {0.077} & {0.292} & {0.075} & {0.164} & {0.324} & {0.529}
  & {0.071} & {0.359} & {0.084} & {0.145} & {0.227} & {0.307} \\

\TypeLLM{Mistral-24b}
  & 0.118 & 0.329 & 0.173 & 0.222 & 0.321 & 0.346
  & 0.046 & 0.231 & 0.090 & 0.119 & 0.162 & 0.368
  & 0.072 & 0.355 & 0.087 & 0.144 & 0.177 & 0.250 \\
\quad+RAG (GPT-text-emd-3-large)
  & {0.184} & {0.487} & {0.185} & {0.247} & {0.469} & {0.506}
  & {0.139} & {0.415} & {0.179} & {0.209} & {0.338} & {0.529}
  & {0.121} & {0.481} & {0.169} & {0.228} & {0.325} & {0.413} \\

\TypeLLM{Qwen3-32b}
  & 0.171 & 0.395 & 0.123 & 0.185 & 0.346 & 0.395
  & 0.061 & 0.215 & 0.119 & 0.149 & 0.250 & 0.353
  & 0.065 & 0.342 & 0.083 & 0.131 & 0.242 & 0.332 \\
\quad+RAG (KaLM-Gemma3-12b-emd)
  & {0.171} & {0.553} & {0.210} & {0.222} & {0.383} & {0.457}
  & {0.108} & {0.462} & {0.209} & {0.269} & {0.456} & {0.544}
  & {0.118} & {0.455} & {0.163} & {0.219} & {0.324} & {0.418} \\

\TypeLLM{GPT-4.1}
  & 0.079 & 0.289 & 0.111 & 0.173 & 0.469 & 0.506
  & 0.046 & 0.139 & 0.090 & 0.149 & 0.588 & 0.750
  & 0.075 & 0.322 & 0.066 & 0.132 & 0.416 & 0.486 \\
\quad+Random
  & {0.131} & {0.276} & {0.111} & {0.173} & {0.556} & {0.617}
  & {0.061} & {0.215} & {0.090} & {0.179} & {0.618} & {0.750}
  & {0.071} & {0.309} & {0.075} & {0.144} & {0.381} & {0.461} \\

\TypeLLM{DeepSeek-V3.2}
  & 0.066 & 0.303 & 0.136 & 0.173 & 0.617 & 0.704
  & 0.015 & 0.061 & 0.075 & 0.119 & 0.588 & 0.706
  & 0.067 & 0.296 & 0.090 & 0.149 & 0.375 & 0.448 \\
\quad+RAG (Qwen3-8b-emd)
  & {0.092} & {0.263} & {0.173} & {0.222} & {0.627} & {0.704}
  & {0.077} & {0.169} & {0.164} & {0.164} & {0.603} & {0.662}
  & {0.072} & {0.342} & {0.147} & {0.217} & {0.524} & {0.595} \\
  \midrule

  \TypeLRM{GPT-o4-mini}
    & 0.197 & 0.618 & {0.333} & {0.543} & 0.568 & 0.654
    & 0.339 & 0.692 & 0.343 & \BestBoth{0.567} & {0.647} & {0.794}
    & {0.261} & {0.697} & 0.363 & 0.404 & {0.513} & 0.614 \\
  \quad+Random
    & {0.210} & {0.671} & 0.321 & 0.506 & {0.605} & {0.654}
    & \BestBoth{0.354} & {0.723} & {0.373} & \BestBoth{0.567} & 0.603 & 0.750
    & 0.252 & 0.692 & {0.388} & {0.424} & 0.511 & {0.620} \\
\TypeLRM{DeepSeek-R3.2}
  & \BestBoth{0.276} & 0.684 & 0.346 & 0.531 & {0.593} & {0.654}
  & 0.246 & \BestBoth{0.754} & {0.328} & 0.492 & {0.603} & 0.735
  & 0.172 & \BestBoth{0.708} & 0.331 & 0.483 & 0.526 & 0.623 \\
\quad+Random
  & 0.263 & \BestBoth{0.697} & \BestBoth{0.395} & \BestBoth{0.630} & 0.588 & 0.630
  & {0.262} & 0.739 & {0.328} & {0.507} & 0.574 & {0.750}
  & 0.169 & 0.692 & 0.333 & 0.479 & 0.518 & 0.623 \\
  \midrule

  \TypeSFT{Gemma3-4b (+LoRA)}
    & 0.237 & 0.513 & 0.333 & 0.420 & 0.432 & 0.457
    & 0.200 & 0.600 & 0.298 & 0.373 & 0.603 & 0.676
    & 0.228 & 0.642 & 0.437 & 0.511 & 0.680 & 0.751 \\
  \TypeSFT{LLama3.1-8b-Instruct (+LoRA)}
    & 0.250 & 0.605 & 0.383 & 0.432 & \BestBoth{0.691} & \BestBoth{0.716}
    & 0.215 & 0.646 & \BestBoth{0.463} & 0.522 & \BestBoth{0.765} & \BestBoth{0.853}
    & \BestBoth{0.270} & 0.678 & \BestBoth{0.574} & 0.642 & \BestBoth{0.730} & \BestBoth{0.787} \\
  \TypeSFT{Qwen3-8b (+LoRA)}
    & 0.263 & 0.632 & 0.358 & 0.444 & 0.531 & 0.556
    & 0.185 & 0.585 & 0.313 & 0.418 & 0.647 & 0.706
    & 0.260 & 0.690 & 0.563 & \BestBoth{0.649} & 0.710 & 0.768 \\

  \bottomrule
\end{tabular}%
}%
} 
\usebox{\MainTabBox}

\end{table*}

We conduct experiments to answer these research questions (\textbf{RQs}).
\begin{itemize}[leftmargin=*, topsep=2pt, itemsep=1pt, parsep=0pt, partopsep=0pt]
\item \textbf{RQ1:} How do existing advanced LLMs perform in the itinerary modification task?
\item \textbf{RQ2:} How do different RAG methods influence LLMs' performance?
\item \textbf{RQ3:} How do SFT paradigms affect LLMs' performance?
\item \textbf{RQ4:} Can SFT and RAG be combined to yield consistent incremental improvements?
\end{itemize}

\subsection{Experimental Setup}
{\bfseries Datasets and Baselines.} 
We utilize the iTIMO for our experiments, evaluating each operation across all sub-datasets to provide a fine-grained analysis. Each dataset is split into training, validation, and test sets in a 7:1:2 ratio. {To conduct comprehensive benchmark across varying \textbf{parameter scales}, \textbf{model families}, \textbf{reasoning capabilities}, and \textbf{accessibilities} (\ie open-weights or proprietary)}, we choose eight base LLMs ({\itshape i.e.,} Gemma-4b {\cite{gemma3_report_2025}}, Llama-3.1-8b {\cite{dubey2024llama}}, Qwen3-8/32b {\cite{qwen3_report_2025}}, Phi4-15b {\cite{phi4_report_2024}}, Mistral-24b {\cite{mistral_small3_2025}}, DeepSeek V3.2 {\cite{deepseek_v3_2_exp_release_2025}}, and GPT-4.1 {\cite{openai-gpt4.1}}), and two Large Reasoning Models (LRMs, {\itshape i.e.,} DeepSeek R3.2 {\cite{deepseek-r1}} and GPT-o4 mini {\cite{openai_o4mini_model_docs}}). To further assess the impact of domain adaptation, we apply supervised fine-tuning (SFT) to {representative} LLMs via Low-Rank Adaptation (LoRA) \cite{hu2022lora}.

\noindent {\bfseries Evaluation Protocols.}
We utilize Modification Accuracy (Mod) and All Pass Rate (APR) for evaluations. Mod serves as a strict metric to verify alignment between modification results and ground-truth, whereas APR acts as a soft metric to assess whether the outputs satisfy the provided hints. For instance, given an {\itshape need-to-modify} itinerary and its hint requiring to modify the popularity distribution, APR equals to 1 if the modification result only changes the popularity distribution and remains the other two distributions invariant. The { proposed} metrics in Section 4 are used for {measuring} whether distribution shifts here.

\noindent {\bfseries Experimental Settings.}
We design prompts including task definitions and structured inputs for experiments. Specifically, the input comprises the {\itshape need-to-modify} itinerary, candidate POIs with 4 negative samples and 1 ground-truth, and a high-level hint specifying the modification { intents} (\eg diversity, popularity or distance). Moreover, we employ three RAG strategies \cite{zhang2024, chang2024comprehensive} (\ie random, sparse, and dense) to select related example for in-context learning. The random strategy randomly samples $K$ training examples, and the sparse strategy {(Hint)} retrieves the top-$K$ by hint similarity. The dense strategy embeds the input as query and { retrieves} top-$K$ similar examples from the training data by cosine similarity. We set $K=3$ and use three retrievers including KaLM-Gemma3-12b-emd \cite{zhao2025kalmv2}, Qwen3-8b-emd \cite{zhang2025qwen3embedding}, and GPT-text-emd-3-large \cite{openai_text_embedding_3_large_docs} due to their strong performance on MMTEB benchmark \cite{enevoldsen2025mmteb}. Please refer to our github repository for more implementation details.

\subsection{Main Results Analysis (RQ1)}
Table~{\ref{tab:main_result}} presents the experimental results yielding the following findings. {{\bf (1)} Current LLMs demonstrate high performance on the single-stage DELETE operation but struggle with multi-stage operations which necessitate both position and {candidate} POI selections. This limitation is exacerbated in ADD. It may be because ADD operation involves one more insertion position than REPLACE operation. This extra option amplifies the decision complexity as the number of candidates grows.} {\bf (2)} Within the Qwen3 family, Qwen3-32b outperforms Qwen3-8b in most cases. This confirms that scaling parameter size effectively enhances model capabilities in handling itinerary modifications. {\bf (3)} {Performance varies significantly across model families even with identical parameter counts.} Despite both being 8B models, Qwen3-8b generally surpasses Llama-3.1-8b across various settings. {\bf (4)} {While RAG improves base LLMs, SFT is particularly notable because it enables 8B models to compete with much larger models {(\eg GPT-o4-mini)}.} {\textbf{In summary, we answer RQ1:} {\itshape Existing LLMs face significant challenges in complex multi-stage modifications. These limitations can be mitigated through model scaling, strategic backbone selection, and targeted enhancements like RAG and SFT, the latter of which allows smaller models to achieve performance comparable with much larger counterparts.}} 

\subsection{In-depth Analysis (RQ2--RQ4)}
{\bfseries Effects of Retriever (RQ2).} Similar to \cite{yin2023understanding}, we adopt a Borda count \cite{emerson2013original} to aggregate our experimental results for better overall comparisons regarding the effects of different retrievers. Specifically, we rank different RAG settings' {(\ie 1 random, 1 sparse, and 3 three dense strategies)} performance for each base LLM, each modification operation, and each dataset. The best setting receives 4 points, the worst setting receives 0 points, and ties are handled by assigning the averaged points of the tied ranks. We compute the Borda count separately for Mod and APR ($8$ base LLMs $\times 3$ datasets $\times 3$ operations $= 72$ cases per metric). Table~\ref{tab:base_rag_borda_mod_apr} reports the experimental results. {\bf Overall, we answer RQ2:} {\itshape RAG significantly enhances performance, with dense retrieval consistently yielding the most substantial improvements compared to other methods.}

\noindent {\bfseries LoRA {vs.} Full Fine-Tuning (RQ3).} Table~\ref{tab:sft_lora_fullft} reports the experimental results regarding the influence of different SFT paradigms. We observe that FullFT generally outperforms LoRA on smaller datasets like Melbourne and Toronto. However, this performance gap narrows on the large-scale Florence dataset. We attribute this observation to the sufficiency of low-rank updates in data-abundant scenarios. With more training instances, LoRA effectively captures the necessary task adaptations without requiring updates to the entire parameter space. {\bf In conclusion, we answer RQ3:} {\itshape FullFT is preferable in data-scarce scenarios for superior adaptability, whereas parameter-efficient LoRA serves as a competitive and resource-efficient alternative as data scale increases.}

\begin{table}[t]
\centering
\caption{Comparison of different RAG settings on base LLMs, where {\texttt{Borda}} denotes the Borda count, \texttt{Rank} is the final order induced by Borda, and $\Delta$ is the mean improvement over the zero-shot setting.}
\label{tab:base_rag_borda_mod_apr}
\footnotesize
\resizebox{\columnwidth}{!}{%
\begin{tabular}{@{}l|ccc|ccc@{}}
\toprule
\multirow{2}{*}{\textbf{RAG Setting}}
& \multicolumn{3}{c|}{\textbf{Modification Accuracy}} & \multicolumn{3}{c}{\textbf{All Pass Rate}} \\
\cmidrule(lr){2-4}\cmidrule(lr){5-7}
& \textbf{Rank}$\downarrow$ & \textbf{Borda}$\uparrow$ & $\Delta\uparrow$
& \textbf{Rank}$\downarrow$ & \textbf{Borda}$\uparrow$ & $\Delta\uparrow$ \\
\midrule
Random
& 5 & 92.0  & +2.77\%
& 5 & 85.5  & +3.48\% \\
\midrule
Hint
& 4 & 102.5 & +2.24\%
& 4 & 106.0 & +3.06\% \\
\midrule
GPT-text-emd-3-large
& \textbf{1} & \textbf{180.0} & +4.42\%
& \textbf{1} & \textbf{180.5} & +5.35\% \\
KaLM-Gemma3-emd
& 3 & 172.0 & +4.30\%
& 3 & 171.0 & +5.24\% \\
Qwen3-8b-emd
& 2 & 173.5 & \textbf{+4.49\%}
& 2 & 177.0 & \textbf{+5.97\%} \\
\bottomrule
\end{tabular}%
}
\end{table}
\definecolor{DeltaPos}{RGB}{210,240,210} 
\definecolor{DeltaNeg}{RGB}{252,214,214} 
\newcommand{\PosDelta}[1]{\cellcolor{DeltaPos}#1}
\newcommand{\NegDelta}[1]{\cellcolor{DeltaNeg}#1}
\begin{table}[t]
\centering
\caption{SFT results (LoRA vs.\ FullFT), where
$\Delta$ denotes the relative average change of FullFT over LoRA, \legendrect{DeltaPos} indicates $\Delta>0$ (improvement), and \legendrect{DeltaNeg} indicates $\Delta\le 0$ (no improvement).}
\label{tab:sft_lora_fullft}
\scriptsize
\setlength{\tabcolsep}{2pt}
\resizebox{\columnwidth}{!}{%
\begin{tabular}{@{}c|c|cc|cc|cc|c@{}}
\toprule
\multirow{2}{*}{\textbf{Methods}} & \multirow{2}{*}{\textbf{Dataset}}
& \multicolumn{2}{c|}{\textbf{ADD}} & \multicolumn{2}{c|}{\textbf{REPLACE}} & \multicolumn{2}{c|}{\textbf{DELETE}} & \multirow{2}{*}{$\Delta\uparrow$} \\
\cmidrule(lr){3-4}\cmidrule(lr){5-6}\cmidrule(lr){7-8}
& & \textbf{Mod}${\uparrow}$ & \textbf{APR}${\uparrow}$
  & \textbf{Mod}${\uparrow}$ & \textbf{APR}${\uparrow}$
  & \textbf{Mod}${\uparrow}$ & \textbf{APR}${\uparrow}$ \\
\midrule
\multirow{3}{*}{\makecell[c]{Gemma3-4b\\(FullFT)}}
  & Melbourne & 0.197 & 0.566 & 0.272 & 0.309 & 0.568 & 0.580 &\PosDelta{+4.2\%} \\
  & Toronto  & 0.292 & 0.662 & 0.463 & 0.552 & 0.632 & 0.662 &\PosDelta{+18.6\%} \\
  & Florence & 0.214 & 0.622 & 0.457 & 0.535 & 0.692 & 0.762 &\PosDelta{+1.0\%} \\
\midrule
\multirow{3}{*}{\makecell[c]{Llama3.1-8b\\(FullFT)}}
  & Melbourne & 0.263 & 0.632 & 0.457 & 0.494 & 0.667 & 0.704 &\PosDelta{+4.5\%} \\
  & Toronto  & 0.277 & 0.631 & 0.508 & 0.567 & 0.647 & 0.765 &\NegDelta{-2.0\%} \\
  & Florence & 0.257 & 0.676 & 0.562 & 0.645 & 0.736 & 0.795 &\NegDelta{-0.3\%} \\
\midrule
\multirow{3}{*}{\makecell[c]{Qwen3-8b\\(FullFT)}}
  & Melbourne & 0.276 & 0.671 & 0.444 & 0.519 & 0.667 & 0.679 &\PosDelta{+17.0\%} \\
  & Toronto  & 0.277 & 0.677 & 0.493 & 0.597 & 0.735 & 0.838 &\PosDelta{+26.8\%} \\
  & Florence & 0.284 & 0.715 & 0.607 & 0.675 & 0.730 & 0.773 &\PosDelta{+3.9\%} \\
\bottomrule
\end{tabular}
}
\end{table}


\definecolor{DeltaPosBg}{RGB}{210,240,210}
\definecolor{DeltaNegBg}{RGB}{245,210,210}

\newcommand{\deltacell}[1]{%
  \begingroup
  \edef\temp{#1}%
  \def\DeltaZeroA{+0.0\%}%
  \def\DeltaZeroB{0.0\%}%
  \ifx\temp\DeltaZeroA
    \cellcolor{DeltaNegBg}#1%
  \else\ifx\temp\DeltaZeroB
    \cellcolor{DeltaNegBg}#1%
  \else
    \expandafter\deltacellaux\temp\relax
  \fi\fi
  \endgroup
}
\def\deltacellaux#1#2\relax{%
  \ifx#1-%
    \cellcolor{DeltaNegBg}#1#2%
  \else
    \cellcolor{DeltaPosBg}#1#2%
  \fi
}

\newcolumntype{M}{c}
\newcolumntype{A}{c}
\newcolumntype{L}[1]{>{\raggedright\arraybackslash}p{#1}}

\begin{table}[t]
  \centering
  \caption{Effects of various RAG settings on LLMs with SFT, where $\Delta$ reports the mean change over the corresponding zero-shot settings.}
  \label{tab:RQ4}

  \scriptsize
  \setlength{\tabcolsep}{1.5pt}

  \begin{tabular}{@{}c|L{0.24\columnwidth}| M A| M A| M A@{\hskip\tabcolsep}}

    \toprule
    \multirow{2}{*}{\textbf{Backbone}} &
    \multicolumn{1}{c|}{\multirow{2}{*}{\makecell[c]{\textbf{RAG Setting}\\\textbf{(Inference)}}}} &
    \multicolumn{2}{c|}{\makecell[c]{\textbf{Zero-shot}\\\textbf{(Train)}}} &
    \multicolumn{2}{c|}{\makecell[c]{\textbf{Random}\\\textbf{(Train)}}} &
    \multicolumn{2}{c}{\makecell[c]{\textbf{Alignment}\\\textbf{(Train)}}} \\
    \cmidrule(lr){3-4}\cmidrule(lr){5-6}\cmidrule(lr){7-8}
    & &
    $\Delta_{\mathrm{Mod}}\uparrow$ & 
    $\Delta_{\mathrm{APR}}\uparrow$ &
    $\Delta_{\mathrm{Mod}}\uparrow$ & 
    $\Delta_{\mathrm{APR}}\uparrow$ &
    $\Delta_{\mathrm{Mod}}\uparrow$ & 
    $\Delta_{\mathrm{APR}}\uparrow$ \\
    \midrule

\multirow{5}{*}{\shortstack[c]{Gemma3-4b\\(LoRA)}} & Random
  & \deltacell{-44.3\%} & \deltacell{-36.6\%} & \deltacell{+5.6\%} & \deltacell{+4.7\%} & \deltacell{+5.6\%} & \deltacell{+4.7\%} \\
& Hint
  & \deltacell{-35.7\%} & \deltacell{-32.3\%} & \deltacell{-1.1\%} & \deltacell{+2.5\%} & \deltacell{+1.7\%} & \deltacell{-4.8\%} \\
& GPT-text-emd-large
  & \deltacell{-35.4\%} & \deltacell{-30.5\%} & \deltacell{+4.3\%} & \deltacell{+6.6\%} & \deltacell{+14.9\%} & \deltacell{+0.6\%} \\
& KaLM-Gemma3-emd
  & \deltacell{-39.1\%} & \deltacell{-31.5\%} & \deltacell{+7.5\%} & \deltacell{+7.2\%} & \deltacell{+13.5\%} & \deltacell{+2.3\%} \\
& Qwen3-8b-emd
  & \deltacell{-38.8\%} & \deltacell{-29.9\%} & \deltacell{+4.8\%} & \deltacell{+5.1\%} & \deltacell{+15.2\%} & \deltacell{+2.6\%} \\
\midrule


\multirow{5}{*}{\shortstack[c]{Llama3.1-8b\\(LoRA)}} & Random
  & \deltacell{-8.2\%} & \deltacell{-7.6\%} & \deltacell{-0.1\%} & \deltacell{+9.5\%} & \deltacell{-0.1\%} & \deltacell{+9.5\%} \\
& Hint
  & \deltacell{-15.1\%} & \deltacell{-6.8\%} & \deltacell{-5.0\%} & \deltacell{+4.8\%} & \deltacell{-9.1\%} & \deltacell{+2.8\%} \\
& GPT-text-emd-large
  & \deltacell{-10.8\%} & \deltacell{-7.5\%} & \deltacell{-2.2\%} & \deltacell{+6.0\%} & \deltacell{+3.8\%} & \deltacell{+3.8\%} \\
& KaLM-Gemma3-emd
  & \deltacell{-10.1\%} & \deltacell{-7.2\%} & \deltacell{-1.2\%} & \deltacell{+5.9\%} & \deltacell{+1.1\%} & \deltacell{+1.3\%} \\
& Qwen3-8b-emd
  & \deltacell{-7.8\%} & \deltacell{-6.3\%} & \deltacell{-4.1\%} & \deltacell{+5.3\%} & \deltacell{+2.9\%} & \deltacell{+5.5\%} \\
\midrule


\multirow{5}{*}{\shortstack[c]{Qwen3-8b\\(LoRA)}} & Random
  & \deltacell{-17.2\%} & \deltacell{-10.3\%} & \deltacell{+7.7\%} & \deltacell{+9.3\%} & \deltacell{+7.7\%} & \deltacell{+9.3\%} \\
& Hint
  & \deltacell{-3.5\%} & \deltacell{-3.4\%} & \deltacell{+9.9\%} & \deltacell{+12.6\%} & \deltacell{+6.2\%} & \deltacell{+9.8\%} \\
& GPT-text-emd-large
  & \deltacell{-3.3\%} & \deltacell{-5.2\%} & \deltacell{+11.3\%} & \deltacell{+9.0\%} & \deltacell{+16.2\%} & \deltacell{+11.1\%} \\
& KaLM-Gemma3-emd
  & \deltacell{-8.7\%} & \deltacell{-3.4\%} & \deltacell{+7.2\%} & \deltacell{+4.1\%} & \deltacell{+18.0\%} & \deltacell{+13.9\%} \\
& Qwen3-8b-emd
  & \deltacell{-4.7\%} & \deltacell{-4.0\%} & \deltacell{+5.8\%} & \deltacell{+8.6\%} & \deltacell{+20.7\%} & \deltacell{+14.8\%} \\


    \bottomrule
  \end{tabular}
\end{table}

\noindent {\bfseries Effects of Incorporating RAG and SFT (RQ4).} 
{Since SFT and RAG individually enhance base LLMs' performance, we investigate their synergistic potential. Specifically, we evaluate the impact of various RAG settings during \textbf{inference (Rows)} across models tuned via three distinct SFT \textbf{training protocols (Columns)}. The results are presented in Table~\ref{tab:RQ4}.
The \textbf{"Zero-shot (Train)"} column reveals that applying RAG inference to models fine-tuned on zero-shot prompts results in negative $\Delta$ values, indicating a degradation compared to standard zero-shot inference. We attribute this to a training-inference mismatch.
To verify this, we introduce \textbf{"Random (Train)"} (fine-tuning with random examples) and \textbf{"Alignment (Train)"} (using the \textit{identical} retrieval setting for both training and inference). The positive $\Delta$ values in these columns confirm that bridging this gap allows SFT models to effectively leverage the retrieved examples.
{\bf Consequently, we answer RQ4:} {\itshape Naively combining SFT and RAG does not always guarantee improvements, and prompt alignment between training and inference is essential.} }

\section{Conclusion and Future Work}
    This paper introduces the critical task of {\itshape itinerary modification} and proposes iTIMO, a dataset constructed via LLM-based perturbation of real-world itineraries. We comprehensively benchmark advanced LLMs on iTIMO, providing a valuable resource for the TRS community. However, our study remains certain limitations that suggests directions for future work. 
Regarding the \textbf{dataset}, we acknowledge limitations regarding user-agnostic generation and absent temporal features due to source inaccuracies. Additionally, the current dataset is constrained in itinerary sequence length and overall scale. 
Future research could apply our proposed pipeline to larger source datasets with longer itineraries, thereby scaling up both the volume and complexity of the generated data. Moreover, incorporating user simulation \cite{wei2025mirroring} and temporal constraints would provide richer contexts (\eg perturbing itineraries via the shuffle operation based on temporal perturbation intents).
\textbf{Methodologically}, our results identify SFT as a promising approach for this task. Subsequent studies might design novel reinforcement learning (RL) or diffusion-based methods to further enhance performance.


\begin{acks}
This work is supported by National Natural Science Foundation of China (71971221 and 62477009), Natural Science Foundation of Hunan Province (2025JJ50419), Major Scientific and Technological Innovation Platform Project of Hunan Province (2024JC1003), Changsha Natural Science Foundation of China (kq2502127), the Ministry of Education, Singapore, under its Academic Research Fund (AcRF) Tier 1 grant, and funded through the SMU-SUTD Internal Research Grant Call (SMU-SUTD 2023\_02\_01), and in part by the Ministry of Education, Singapore, under its Academic Research Fund Tier 2 (Award No. MOE-T2EP2012 30015).
\end{acks}

\bibliographystyle{ACM-Reference-Format}
\balance
\bibliography{sample-base}

\end{document}